\documentclass[twocolumn,showpacs,amsmath,amssymb,floatfix,superscriptaddress]{revtex4}
\usepackage{graphicx}
\usepackage{dcolumn}
\usepackage{bm}
\usepackage{hyperref} 
\usepackage{latexsym}

\begin{document}
\title{Kinetic Anomalies in Addition-Aggregation Processes} 
\author{M. Mobilia}
\email{mmobilia@buphy.bu.edu} \author{P.~L.~Krapivsky} \email{paulk@bu.edu}
\author{S.~Redner} \email{redner@bu.edu} \affiliation{Center for BioDynamics,
  Center for Polymer Studies, and Department of Physics, Boston University,
  Boston, MA, 02215}

\begin{abstract}
  We investigate irreversible aggregation in which monomer-monomer,
  monomer-cluster, and cluster-cluster reactions occur with constant but
  distinct rates $K_{\rm MM}$, $K_{\rm MC}$, and $K_{\rm CC}$, respectively.
  The dynamics crucially depends on the ratio $\gamma=K_{\rm CC}/K_{\rm MC}$
  and secondarily on $\epsilon=K_{\rm MM}/K_{\rm MC}$.  For $\epsilon=0$ and
  $\gamma<2$, there is conventional scaling in the long-time limit, with a
  single mass scale that grows linearly in time.  For $\gamma\ge 2$, there is
  unusual behavior in which the concentration of clusters of mass $k$, $c_k$
  decays as a stretched exponential in time within a boundary layer
  $k<k^*\propto t^{1-2/\gamma}$ ($k^*\propto \ln t$ for $\gamma=2$), while
  $c_k\propto t^{-2}$ in the bulk region $k>k^*$.  When $\epsilon>0$,
  analogous behaviors emerge for $\gamma<2$ and $\gamma\geq 2$.
\end{abstract}

\maketitle \date{\today}

\section{Introduction}

In this work, we investigate a simple aggregation process in which three
kinds of reactions occur with constant but distinct rates:
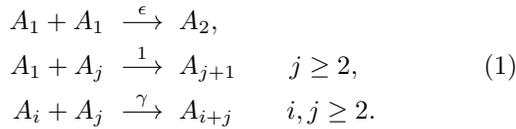
\begin{eqnarray}
\label{AA}
A_{1}+A_{1}& \stackrel{\epsilon}{\longrightarrow} & A_{2}, \nonumber\\
A_{1}+A_{j}& \stackrel{1}{\longrightarrow} & A_{j+1}\qquad j\geq 2, \\
A_{i}+A_{j}&\stackrel{\gamma}{\longrightarrow}& A_{i+j}\qquad 
i,j\geq 2.\nonumber
\end{eqnarray}
Equivalently, the reaction matrix $K_{ij}$ has the value $\epsilon$ in the
upper left corner, the value 1 for all elements along the top and left edges
of the matrix, and the value $\gamma$ in the rest of the matrix.

Despite its simplicity, this model exhibits rich dynamics in which both
scaling and universality can be violated.  The model was initially studied by
Hendriks and Ernst \cite{HE} to describe polymerization in which addition
processes (reactions involving monomers) occur more readily than aggregation
(reactions between $i$ and $j$, with $i,j \geq 2$).  They found evidence of
unusual kinetic behavior by an implicit solution for the cluster
concentrations; see also \cite{SP} for related work.  Our focus is on the
opposite limit in which aggregation dominates over addition.  By explicit
solutions of the rate equations, we elucidate the full range of behaviors for
this system.

{}From a general perspective, the origin of the anomalous behavior in
addition/aggregation stems from the fact that the simplest case of
aggregation with a constant reaction rate is actually a marginal system.  As
discussed by van Dongen and Ernst \cite{DE}, aggregation can be broadly
categorized by whether reactions among large clusters dominate (type I
systems, in their nomenclature) or whether reactions between large and small
clusters dominate (type III systems).  The former generically leads to
monotonically decaying cluster mass distributions, while the latter gives
peaked distributions.  The constant reaction rate system (type II system) is
marginal by being at the boundary between these two behaviors.

Because of this marginality, the kinetics of constant-kernel aggregation is
sensitive to small perturbations in the reaction rates.  Previously-studied
examples of this feature include aggregation involving two distinct monomeric
units, so that the reaction rates between two even masses, two odd masses, or
an even and an odd mass are naturally distinct.  Here non-universal and
non-scaling behavior arises as a function of these rates \cite{LR,CL}.
Another example is aggregation with the rate $K_{ij}=2-q^i-q^j$ with $0<q<1$
\cite{CLq}.  Although this reduces to a constant-kernel system for
$i,j\to\infty$, the $q$-dependent terms lead to unusual kinetics.  The model
studied here is in the spirit of the first example, except that we perturb
only an infinitesimal fraction of the reaction rates in a constant-kernel
system.  It is striking that such a small change in the reaction rates has
such a profound influence on the kinetics.

In the next two sections, we focus on the special case of ``sterile''
monomers, where monomer-monomer reactions do not occur.  In Sec.~II, we first
determine the monomer and dimer concentrations and show that different
behaviors arise for $\gamma\geq 2$ and $\gamma<2$.  Then in Sec.~III, we
derive asymptotic results for the cluster concentrations.  We then study, in
Sec.~IV, these same three cases when monomer-monomer reactions can also
occur.  Sec.~V gives a summary as well as a discussion of the equivalence
between the model and a diffusion-controlled process in which monomers have
different diffusivity than other clusters.

\section{Sterile Monomers} 

When monomers do not interact among themselves ($K_{11}=0$), we shall show
that the system exhibits three distinct kinetic regimes, with unusual time
dependences and breakdown of conventional scaling in two of these cases.

The rate equation for the density $c_{k}(t)$ is \cite{RE}
\begin{eqnarray}
\label{ckt}
\dot c_k(t)=\frac{1}{2}\sum_{i+j=k}K_{ij}c_{i}(t)c_{j}(t)-
c_{k}(t)\sum_{j\geq 1}K_{kj}c_{j}(t),
\end{eqnarray}
where the overdot denotes time derivative.  The first (gain) term accounts
for the formation of clusters of mass $k$ ($k$-mers) as a result of the
bimolecular aggregation of $i$-mers with $j$-mers, with $i+j=k$.  The second
(loss) term accounts for the aggregation of the $k$-mers with any $j$-mer,
thus leading to a decrease in the $k$-mer concentration.  For the processes
given in (\ref{AA}) and with the additional constraint of
$K_{11}=\epsilon=0$, the explicit rate equations are
\begin{eqnarray}
\label{RE}
\dot c_1&=&-c_{1} R,\nonumber\\
\dot c_2&=&-c_{2}\left[c_1 +\gamma R\right]\\
\dot c_k&=& c_1c_{k-1}+\frac{\gamma}{2}
\sum_{i+j=k}c_{i}c_{j}-c_{k} (c_{1}+\gamma R)\quad k>2,\nonumber
\end{eqnarray}
where we drop the time argument for compactness and introduce the shorthand
$R\equiv \sum_{k\geq 2} c_{k}$ for the concentration of reactive clusters
(those with mass $\geq 2$).  Notice also that there is no production of
monomers and dimers in the reaction; they can only disappear.

The governing equation for $R$ is obtained by summing Eqs.~(\ref{RE}) to give
$\dot R=-\frac{\gamma}{2}R^{2}$, whose solution is
\begin{eqnarray}
\label{Rt}
R=R(0)\,\tau^{-1}, \qquad \tau\equiv 1+\frac{1}{2}\gamma R(0)\,t.
\end{eqnarray}
With the explicit expression for $R$, we integrate the equation for monomers
to give
\begin{eqnarray}
\label{c1}
c_1=c_{1}(0)\,\tau^{-2/\gamma}.
\end{eqnarray}
The crucial point is that the value of $\gamma$ controls which of $R$ or
$c_1$ decays more rapidly.  For $\gamma<2$, monomers are asymptotically
irrelevant and conventional constant-kernel aggregation kinetics arises.  For
$\gamma\geq 2$, however, the large disparity in the reaction rates $K_{1j}$
and $K_{ij}$ leads to a long-lived residue of monomers and these profoundly
affect the asymptotic kinetics.

To see this explicitly, we solve for the density of dimers to find
\begin{eqnarray}
\label{c2}
c_2= c_2(0)\,\tau^{-2}\,e^{-\lambda}\,,
\end{eqnarray}
where we define $\lambda=\int_0^t dt'c_1(t')$.  Using Eq.~(\ref{c1}) we
express $\lambda$ in terms of $\tau$:
\begin{equation}
\label{lambda}
\lambda\equiv \lambda(\tau)=\left\{ \begin{array}{ll}
2 r\,\frac{\tau^{1-2/\gamma} -1}{\gamma -2} &\quad\gamma\ne 2, \\ \\ 
r\,\ln\tau  & \quad\gamma=2.
\end{array} 
\right. 
\end{equation}
where we further define $r=c_1(0)/R(0)$.

Depending of the value of $\gamma$ we obtain three distinct long-time
behaviors:
\begin{itemize}
  
\item For $0<\gamma< 2$, $\lambda\rightarrow \frac{2r}{2-\gamma}\to{\rm
    const}$ as $t\to\infty$.  Thus the dimer concentration has a universal
  $t^{-2}$ asymptotic decay; only the amplitude of the decay depends on the
  initial state.
 
\item For $\gamma> 2$, the dimer density decays as a stretched exponential
  ${\rm e}^{-\lambda}$ times a power-law prefactor.
  
\item For $\gamma=2$, the dimer concentration asymptotically decays as
  $t^{-(2+r)}$, where the exponent depends on the initial concentration ratio
  $r=c_1(0)/R(0)$.
\end{itemize}

\section{Asymptotic Mass Distribution}

To determine the asymptotic cluster concentrations $c_k$, we re-write the
rate equations (\ref{RE}) in the form
\begin{eqnarray}
\label{cknew}
\dot{c}_k+\alpha c_k=\alpha_k,
\end{eqnarray}
where
\begin{eqnarray}
\label{alpha} 
\alpha\equiv c_1+\gamma R, \quad
\alpha_{k}\equiv c_1 c_{k-1}+
\frac{\gamma}{2}\sum_{\stackrel{i+j=k}{i,j\geq 2}}c_{i}c_{j}.
\end{eqnarray}
Due to the recursive nature of Eq.~(\ref{cknew}) we immediately obtain the
formal solution for $c_{k}$:
\begin{eqnarray}
\label{cksol}
c_{k}(t)={\cal E}(t)\left\{c_{k}(0)+
\int_{0}^{t} dt'\, \frac{\alpha_{k}(t')}{{\cal E}(t')}\right\}
\end{eqnarray}
with ${\cal E}(t)=\exp\left[-\int_{0}^{t}dt'
  \alpha(t')\right]=\tau^{-2}\,e^{-\lambda}$.  Therefore, once we know
$c_1,\dots, c_{k-1}$, we can compute $\alpha_k$ and then determine $c_k$.

Before proceeding to general $k$, it is instructive to compute $c_{3}$.
{}From Eq.~(\ref{cksol}), and using the previous results for $c_1$ and $c_2$,
we obtain:
\begin{eqnarray}
\label{c3}
c_3=\left[c_2(0)\,\lambda +c_3(0)\right]\,\tau^{-2}\,e^{-\lambda}\;.
\end{eqnarray}
As in the case of dimers, there are three distinct asymptotic behaviors for
the trimer concentration that are best expressed in terms of $c_3/c_2$.  We
obtain
\begin{eqnarray}
\label{c32}
\frac{c_3}{c_2}\propto \left\{\begin{array}{ll}
  {\displaystyle t^{1-2/\gamma}}&\quad \gamma>2, \\  
  {\displaystyle \ln t}         &\quad \gamma=2, \\  
  {\displaystyle 1}             &\quad \gamma<2, 
\end{array} 
\right. 
\end{eqnarray}
that is, $c_3/c_2\propto \lambda$ for any value of $\gamma$ (see
Eq.~(\ref{lambda}).  While the exact expressions for the concentrations $c_k$
become unwieldy as $k$ grows, the asymptotic expressions turn out to be
simple: $c_3/c_2\to\lambda$, $c_4/c_2\to\lambda^2/2$, {\it etc}.  We will
confirm this hypothesis by induction in the following subsections.

\subsection{$\gamma\ge 2$}

When $\gamma$ is strictly greater than 2, we easily find that $c_k/c_2\propto
\lambda^{k-2}$ as $t\to\infty$ by explicit calculation for small $k$.  We
therefore seek an asymptotic solution for $c_k$ of the form
\begin{eqnarray}
\label{large}
\frac{c_k}{c_2}=\beta_k\lambda^{k-2} 
+ \ldots
\end{eqnarray}
where $\ldots$ denotes subdominant terms in the limit $t\rightarrow \infty$
and the amplitude $\beta_{k}$ will be determined below.

For concreteness and simplicity, let us consider the bi-disperse initial
condition in which only $c_1(0)$ and $c_2(0)$ are non-zero while $c_{k}(0)=0$
for all $k>2$.  To extract the asymptotics of $c_k$ from Eq.~(\ref{cksol}),
only the second term on the right-hand side is needed.  Let us suppose that
Eq.~(\ref{large}) holds for $c_2,\ldots, c_{k}$.  To leading order, we have
$\alpha_{k+1}=c_1c_k$, and substituting $c_k=c_2\beta_k\lambda^{k-2}$ into
Eq.~(\ref{cksol}), we obtain
\begin{eqnarray}
\label{k+1}
c_{k+1}
&=&{\cal E}(t)
\int_0^t dt'\,\frac{c_{1}(t')\, c_k(t')}{{\cal E}(t')}\nonumber\\
&=&c_2\,\,\frac{\beta_k}{k-1}\,
\lambda^{k-1}\,, 
\end{eqnarray}
where to obtain the second line we use ${\cal E}(t)=\tau^{-2}\,
e^{-\lambda}$.  Thus $\beta_{k}=1/(k-2)!$ and we arrive at the remarkably
simple asymptotic solution
\begin{eqnarray}
\label{ck1}
c_{k+2}=c_2\,\,\frac{\lambda^k}{k!},\qquad
c_2=c_2(0)\,\tau^{-2}\,e^{-\lambda}\,,
\end{eqnarray}
with $\tau$ and $\lambda$ given by Eqs.~(\ref{Rt}) and (\ref{lambda}).  
Explicitly, we have
\begin{equation}
\label{ck-exp}
c_{k+2}\sim {1\over k!}\,\,t^{-2+k(1-2/\gamma)}\,\, 
\exp\big(-{\rm const.}\times t^{1-2/\gamma}\big).
\end{equation}
Thus $c_k$ has a very different asymptotic behavior than in the
constant-kernel system \cite{RE}.  As discussed previously, this anomaly
arises because monomers are very long-lived and these then strongly influence
the long-time kinetics.

One important caveat, however, is that the derivation of Eq.~(\ref{ck1})
applies only for finite $k$.  For sufficiently large $k$ the neglected
subdominant terms accumulate to provide a relevant contribution.  The
simplest way to see that Eq.~(\ref{ck1}) cannot hold over entire mass range
is to use this equation to compute the total cluster density.  This gives
$\sum_{k\geq 2}c_k=c_2\,e^{\lambda}=c_2(0)\,\tau^{-2}$, in contrast to the
correct result $R(0)\tau^{-1}$ given in Eq.~(\ref{Rt}).  

We therefore conclude that the mass distribution naturally divides into a
small-mass boundary layer, $k\leq k^*$, that contains an asymptotically
negligible fraction of the total mass, and a remaining bulk region.  The
extent of the boundary layer may be determined as the maximum of $c_k$ in
Eq.~(\ref{ck1}) and gives $k^*=\lambda\propto t^{1-2/\gamma}$.  Within this
boundary layer, the mass distribution grows rapidly with mass, but this
region contains only an asymptotically negligible fraction of the total mass.
On the other hand, most of the mass, which exhibits conventional scaling
behavior is contained in the bulk region.

In the marginal case of $\gamma=2$, the asymptotic solution of
Eq.~(\ref{ck1}) is still valid.  Since now $\lambda=r\,\ln\tau$, the density
of dimers decays algebraically in time as $c_2=c_2(0)\,\tau^{-(2+r)}$, and
the width of the boundary layer grows logarithmically $k^*=\lambda=r \ln
\tau$.  The following main part of the mass distribution is again
characterized by conventional scaling behavior (see Eq.~(\ref{ckscal})
below). A justification of this picture is given in Appendix.  Thus for
$\gamma\geq 2$, the mass distribution exhibits two growing scales and
conventional single-mass scaling is violated.

\subsection{$\gamma<2$}

By solving $c_2,c_3,\ldots$ exactly, we are led to the hypothesis that for
$k\geq 2$
\begin{eqnarray}
\label{small}
c_{k}\propto \tau^{-2}.
\end{eqnarray}
This indeed can be directly checked by induction.  Moreover, in the scaling
limit,
\begin{eqnarray}
\label{scal}
k\to\infty, \quad t\to\infty, \quad \frac{k}{t}={\rm finite},
\end{eqnarray}
the mass distribution admits the conventional scaling form
\begin{eqnarray}
\label{ckscal}
c_{k}=\frac{4}{\gamma^2t^2}\,e^{-2k/(\gamma t)}.
\end{eqnarray}
Summing $\sum c_k$ we indeed recover $R=2/(\gamma t)$, while the next moment
$\sum kc_k$ equals one (based on the total mass set equal to one initially).
Equation (\ref{ckscal}) also describes the scaling portion of the mass
distribution when $\gamma\geq 2$.

The origin of the conventional scaling behavior for the case $\gamma<2$ is
simple.  For $\gamma<2$ monomers disappear quickly, since $c_1\propto
t^{-2/\gamma}$, and their asymptotic influence is negligible.  Thus the
reaction effectively reduces to a constant-kernel system that begins with
dimers.  The monomers do influence the small-mass behavior, namely $c_k=A_k
t^{-2}$ with mass-dependent amplitudes $A_k$; {\it e.g.}, $A_2=c_2(0)[\gamma
R(0)/2]^{-2}e^{-\lambda}$ and $A_3=A_2[\lambda+c_3(0)/c_2(0)]$.  However, as
$k$ grows the amplitude $A_k$ approaches the constant value $4\gamma^{-2}$.

The existence of scaling can be proved rigorously, {\it e.g.}, by the
generating function approach given in the Appendix.  A simpler approach is to
merely assume that scaling holds and check its consistency {\it a
  posteriori}.  In the continuum limit, we have checked the correctness of
Eq.~(\ref{ckscal}), which is the conventional scaling form for constant
kernel aggregation \cite{RE}.

\section{Reactive monomers}

We now assume that monomer-monomer reactions do occur: $K_{11}\equiv \epsilon
>0$.  This situation has already been studied by Hendriks and Ernst
\cite{DE}.  The rate equations for monomers and dimers now read
\begin{eqnarray}
\label{Rc1}
\dot{c}_1&=&-c_1(\epsilon c_1+R) \\
\label{Rc2}
\dot{c}_2&=&\frac{\epsilon}{2}c_1^2-c_2(c_1 +\gamma R),
\end{eqnarray}
while the rate equations for clusters with $k>2$ are the same as in
Eq.~(\ref{RE}).  Correspondingly, the density of reactive clusters $R$
evolves according to
\begin{equation}
\label{RR}
\dot R=\frac{1}{2}\epsilon c_1^2-\frac{1}{2}\gamma R^2.
\end{equation}

It does not seem possible to solve the coupled non-linear equations
(\ref{Rc1}) and (\ref{RR}) for $c_1$ and $R$ explicitly. If we treat $c_1$ as
a function of $R$, however, we can reduce Eq.~(\ref{Rc1}) and (\ref{RR}) to a
single differential equation that can be solved to establish a functional
relation between $c_1$ and $R$ \cite{HE}.  {}From this relation, we can
deduce that different behaviors emerge depending on whether $\gamma<2$ or
$\gamma\geq 2$.

A simpler and more fruitful way to proceed is to seek the asymptotic behavior
of $c_1$ and $R$ without explicitly solving the governing equations.  We
anticipate that there are three possible asymptotic behaviors: (i) $c_1\ll
R$, (ii) $c_1\propto R$, and (iii) $c_1\gg R$.  Substituting each of these
scenarios into Eqs.~(\ref{Rc1}) and (\ref{RR}) shows that the latter case is
impossible.  Let us now analyze the first two possibilities.  When $c_1\ll
R$, Eq.~(\ref{RR}) gives $R\simeq 2/(\gamma t)$, and then Eq.~(\ref{Rc1})
yields $c_1 \propto t^{-2/\gamma}$.  Thus the relation $c_1\ll R$ holds when
$\gamma<2$.  In the complementary regime of $\gamma>2$, we find that
$c_1\propto R$ is consistent; furthermore, Eqs.~(\ref{Rc1}) and (\ref{RR})
now give $R\propto c_1\propto t^{-1}$.

We now analyze these two cases in more detail.

\subsection{$\gamma\geq 2$}

When $\gamma$ is strictly greater than 2, we substitute the ansatz
\begin{eqnarray}
\label{ansatz}
R\simeq At^{-1}, \qquad c_1 \simeq Bt^{-1},
\end{eqnarray}
into Eqs.~(\ref{Rc1}) and (\ref{RR}) to get a quadratic equation for the
amplitude $A$.  The physical requirement that $A$ and $B$ are both positive
fixes the solution to be
\begin{eqnarray}
\label{AB}
A=\frac{\epsilon -1 +\sqrt{\epsilon(\epsilon+\gamma -2)}}
{\epsilon \gamma-1}, \quad
B=\frac{1-A}{\epsilon},
\end{eqnarray}
The singularity at $\epsilon=\gamma^{-1}$ is only apparent and may be
resolved by applying the l'Hospital's rule, to give $A=\gamma/[2(\gamma-1)]$
in this case.

From the formal solution (\ref{cksol}) we can check by induction that for
finite $k$,
\begin{eqnarray}
\label{ck.1}
 c_{k}\simeq B_kt^{-1}.
\end{eqnarray} 
The amplitude $B_k$ is found by substituting the ansatz (\ref{ck.1}) into
(\ref{cknew})--(\ref{alpha}) to give the recursion
\begin{eqnarray}
\label{BBk}
(\mu-1)B_{k}=BB_{k-1} 
+\frac{\gamma}{2}\sum_{\stackrel{i+j=k}{i,j\geq 2}}B_{i}B_{j}
\end{eqnarray} 
for $k>2$, where we define
\begin{eqnarray}
\label{mu}
\mu=\gamma A+B=1+\sqrt{1+(\gamma-2)/\epsilon}\,.
\end{eqnarray} 
Note also that $(\mu-1)B_{2}=\epsilon B^2/2$.  To solve the recursion
(\ref{BBk}) we introduce the generating function ${\cal B}(z)=\sum_{k\geq 2}
B_k z^k$ to reduce (\ref{BBk}) to
\begin{eqnarray*}
{\cal B}(z)=\frac{\mu-1-Bz}{\gamma}
\left\{1-\sqrt{1-\epsilon\gamma\left(\frac{Bz}{\mu-1-Bz}\right)^2}\right\}\,.
\end{eqnarray*} 
When $k\gg 1$ but still within the boundary layer, the asymptotic behavior of
${\cal B}(z)$ leads to the amplitude
\begin{eqnarray}
\label{Bksol}
B_{k}\simeq C\,k^{-3/2}\beta^k
\end{eqnarray} 
with
\begin{eqnarray}
\label{BC}
\beta=B\,\frac{1+\sqrt{\epsilon\gamma}}{\mu-1}\,,\quad 
C=\frac{\mu-1}{\gamma\sqrt{2\pi}}\,
\sqrt{\frac{\sqrt{\epsilon\gamma}}{1+\sqrt{\epsilon\gamma}}}\,. 
\end{eqnarray} 
One can verify that $\beta<1$ for all $\gamma>2$ and $\epsilon>0$; therefore,
apart from a power-law prefactor, the amplitude $B_k$ decreases exponentially
with $k$.

As in the previous section, the results (\ref{ck.1}) and (\ref{Bksol}) are
valid only within the boundary layer.  Thus the total density of clusters
within the boundary layer is asymptotically equal to $A' t^{-1}$, where
$A'=\sum_{k\geq 2}B_k= {\cal B}(z=1)$, or
\begin{eqnarray}
\label{A'}
A'=\frac{\mu-1-B-\sqrt{(\mu-1-B)^2-\epsilon\gamma B^2}}{\gamma}\,. 
\end{eqnarray} 
Since the total density of clusters scales as $A t^{-1}$, only a fraction
$A'/A$ of clusters lies within the boundary layer.  Using Eqs.~(\ref{AB}) and
(\ref{A'}), we find $A-A'=2/\gamma$; this implies that the density of
clusters in the bulk decays as $2/(\gamma t)$.  The same holds for
$\gamma\leq 2$, although in this case the density of clusters in the boundary
layer is asymptotically negligible.

To determine the asymptotic behavior in the bulk ($k\rightarrow\infty$ and
$t\rightarrow \infty$), we simplify the rate equations by neglecting the
subdominant terms $c_1(c_{k-1}-c_k)$ on the right-hand side of
Eq.~(\ref{RE}).  Further, the sum in (\ref{RE}) has three contributions.
When $i$ lies within the boundary layer, we can replace $\sum_i c_i c_{k-i}$
by $c_k\sum_i B_i t^{-1}=c_k A't^{-1}$; a similar contribution arises when
$j$ lies within the boundary layer.  In the bulk we replace the sum by the
integral. Thus
\begin{eqnarray*}
\frac{\gamma}{2}\sum_{i+j=k}c_{i}c_{j}
\to \frac{\gamma}{2} \int di\,c_{i}c_{k-i}+\frac{\gamma A'}{t}\,c_k\,.
\end{eqnarray*}
Combining all terms and using the identity $\gamma(A-A')=2$ we finally
convert Eq.~(\ref{RE}) to
\begin{eqnarray}
\label{cknewest}
\dot{c}_k+\frac{2}{t}\, c_k=\frac{\gamma}{2}
\int di\,c_{i}c_{k-i}\,.
\end{eqnarray}
This equation appears in the standard constant-kernel aggregation and its
solution, satisfying the aforementioned conservation laws $\int
dk\,c_k=2/\gamma$ and $\int dk\,kc_k=1$, is given by Eq.~(\ref{ckscal}).

By matching the mass distribution in the boundary layer ($c_k\propto
t^{-1}\beta^k$) and in the bulk ($c_k\propto t^{-2}$), we estimate the width
of the boundary layer as
\begin{eqnarray}
\label{k*}
k^*\approx \frac{\ln t}{\ln (1/\beta)}\,,
\end{eqnarray} 
with $\beta$ given by Eq.~(\ref{BC}). This is a much slower growth than in the
case of sterile monomers where $k^*\propto t^{1-2/\gamma}$.

When $\gamma=2$, we again expect marginal behavior.  Proceeding as in the
case of $\gamma>2$, we find
\begin{eqnarray}
\label{Rcmar}
R\simeq t^{-1}, \qquad c_1 \simeq t^{-1}(\epsilon \ln t)^{-1}.
\end{eqnarray}
{}From the formal solution (\ref{cksol}) we can check by induction that,
\begin{eqnarray}
\label{ckmar}
 c_{k}\simeq B_kt^{-1}(\epsilon \ln t)^{-k}.
\end{eqnarray} 
For $k>2$, the amplitude $B_k$ is found from the recursion (\ref{BBk}) with
$\mu=2$, $B=1$, and $B_2=\epsilon/2$.  The generating function and the
coefficients $B_k$ and $C$ are given by taking the corresponding formulae in
the $\gamma>2$ case and setting $\gamma=2$.

This behavior holds in a boundary layer whose width now grows as
\begin{eqnarray}
\label{k*}
k^*\approx \frac{\ln t}{\ln(\Lambda\ln t)}\,, \quad 
\Lambda=\frac{\epsilon}{1+\sqrt{2\epsilon}}\,,
\end{eqnarray} 
while in the bulk we recover the ordinary scaling mass distribution
$c_k=t^{-2}e^{-k/t}$.

\subsection{$\gamma<2$ }

We already argued that $R\simeq 2/(\gamma t)$ and $c_{1}\propto
t^{-2/\gamma}$ for $\gamma<2$.  Solving then for dimers we obtain:
\begin{eqnarray}
\label{c2.2}
{c_2} \propto \left\{\begin{array}{lll}
  {\displaystyle t^{-2}}&\quad 0<\gamma<4/3, \\  
  {\displaystyle t^{1-4/\gamma}}  
&\quad 4/3<\gamma<2.
\end{array} 
\right. 
\end{eqnarray}
Generally, we find
\begin{eqnarray}
\label{ck.2}
{c_k} \propto \left\{\begin{array}{lll}  
{\displaystyle t^{-2}}&\quad 0<\gamma<\frac{2k}{k+1}, \\  \\
{\displaystyle t^{-2}\ln t}&\quad \gamma=\frac{2k}{k+1}, \\  \\
  {\displaystyle t^{k-1-2k/\gamma}} &\quad \frac{2k}{k+1}<\gamma<2.
\end{array} 
\right. 
\end{eqnarray}
Thus for every $\gamma<2$, small-mass clusters have abnormal kinetics: only
monomers for $\gamma<4/3$, monomers and dimers for $4/3\leq \gamma<3/2$,
monomers, dimers, and trimers for $3/2\leq \gamma<8/5$, {\it etc}.  The
remaining clusters decay as $t^{-2}$, and conventional scaling describes the
cluster-mass distribution.

\section{Summary and Discussion} 

For irreversible aggregation with distinct monomer-monomer ($K_{\rm MM}$),
monomer-cluster ($K_{\rm MC}$), and cluster-cluster ($K_{\rm CC}$) reaction
rates, the dynamics depends crucially on the ratio $\gamma=K_{\rm CC}/K_{\rm
  MC}$, while $\epsilon=K_{\rm MM}/K_{\rm MC}$ plays a lesser role --- all
that matters is whether $\epsilon=0$ or $\epsilon>0$.  For $\epsilon=0$ and
$\gamma<2$, there is conventional scaling with a single mass scale growing
linearly with time.  For $\gamma\geq 2$, there are two scales: the boundary
layer $k<k^*\propto t^{1-2/\gamma}$ where the mass distribution has an
unusual Poisson form, and the bulk region $k>k^*$ where conventional scaling
holds.  When $\epsilon>0$ and $\gamma<2$, there is conventional scaling,
except that light clusters have abnormal kinetics -- monomers for
$\gamma<4/3$; monomers and dimers for $4/3\leq \gamma<3/2$; monomers, dimers,
and trimers for $3/2\leq \gamma<8/5$, {\it etc}.  When $\epsilon>0$ and
$\gamma\geq 2$, the behavior in the boundary layer is very different from
that when $\epsilon=0$.  In particular, the mass distribution decays with
mass while for $\epsilon=0$ the mass distribution is peaked.

A possible application of this model is to clustering on surfaces.  Consider
a two-dimensional substrate with diffusing single-layer islands that
aggregate whenever they meet.  In the diffusion-controlled limit, the
reaction rate of an island of radius $R$ and diffusivity $D$ is proportional
to $D\ln R$ \cite{smol}.  To a good approximation, we can ignore the island
radius and think of point-like islands that always occupy a single lattice
site.  When such a cluster hops onto already occupied site, two clusters
immediately coalesce into a single cluster.

If monomers hop with rate $D$ while all heavier clusters hop with the same
unit rate, then the corresponding reaction rates for point-like islands are
$K_{11}=2D, K_{1j}=1+D, K_{ij}=2$.  As a result
\begin{eqnarray}
\label{eg}
\epsilon=\frac{K_{11}}{K_{1j}}=\frac{2D}{1+D}, \quad
\gamma=\frac{K_{ij}}{K_{1j}}=\frac{2}{1+D}. 
\end{eqnarray}

For immobile monomers, $D=0$, we have $\epsilon=0$ and $\gamma=2$.  In this
case, the cluster mass distribution is
\begin{eqnarray}
\label{ckbound}
c_{k+2}=c_2(0)\,\tau^{-(2+r)}\,\frac{\lambda^k}{k!}\,.
\end{eqnarray}
within a boundary layer that grows logarithmically with time $k^*=r\ln
(1+R(0)t)$, where $r={c_1(0)}/{R(0)}$.  In the bulk, ordinary scaling holds
in which $c_k=t^{-2}e^{-k/t}$.

When $D>0$, $\gamma$ is always less than 2.  Therefore,
\begin{eqnarray}
\label{Rc1D}
R=\frac{1+D}{t}, \qquad c_1\propto t^{-1-D}.
\end{eqnarray}
The decay rate for clusters of mass $k>1$ is also simple:
\begin{eqnarray}
\label{ckD}
{c_k} \propto \left\{\begin{array}{lll}
  {\displaystyle t^{-2}}       &\quad D>1/k, \\  
  {\displaystyle t^{-2}\,\ln t}&\quad D=1/k, \\  
  {\displaystyle t^{-1-kD}}    &\quad D<1/k.
\end{array} 
\right. 
\end{eqnarray}

\section{Acknowledgements}
MM thanks the Swiss NSF for financial support under the fellowship
81EL-68473.  PLK and SR thank NSF grant DMR9978902 for financial support of
this research.

\appendix
\section{Generating function approach} 

In the case of sterile monomers, we can employ the generating function
approach as an alternative to a direct solution.  The generating function
\begin{eqnarray}
\label{GF}
G(z,t)\equiv \sum_{k\geq 2}z^{k}c_{k}(t) 
\end{eqnarray}
recasts an infinite set of rate equations into a single differential
equation
\begin{eqnarray}
\label{Gzt}
\frac{\partial G}{\partial t}=\frac{\gamma}{2}G^2+\left(zc_1-\alpha\right) G.
\end{eqnarray}
This is a Bernoulli equation that is readily solved in terms of $G^{-1}$.
The solution is
\begin{eqnarray}
\label{Gsol}
G^{-1}=\frac{\tau^2\,e^{(1-z)\lambda}}{G(z,0)}-
\int_1^\tau \frac{d\tau'}{R(0)}\left(\frac{\tau}{\tau'}\right)^2
e^{(1-z)(\lambda-\lambda')},
\end{eqnarray}
where $\lambda=\lambda(\tau)$ and $\lambda'=\lambda(\tau')$.  For the
bi-disperse initial condition,
$c_k(0)=c_1(0)\delta_{k,1}+c_2(0)\delta_{k,2}$, we have $G(z,0)=c_2(0)\,z^2$
and $R(0)=c_2(0)$.  Using these relations together with
$c_2=c_2(0)\tau^{-2}e^{-\lambda}$ we re-write (\ref{Gsol}) as
\begin{eqnarray}
\label{Gsimple}
G^{-1}=\frac{e^{-z\lambda}}{c_2\,z^2}-\frac{e^{-z\lambda}}{c_2}
\int_1^\tau \frac{d\tau'}{(\tau')^2}\,
e^{-(1-z)\lambda'}\,.
\end{eqnarray}

We now consider in detail the marginal case of $\gamma=2$ and justify the
two-scale structure of the mass distribution.  When $\gamma=2$, we have
$\lambda=r\ln \tau$ and (\ref{Gsimple}) becomes 
\begin{eqnarray}
\label{G2}
G(z,t)=\frac{ c_2(t)\,z^2\,e^{\lambda z} \left(1-\frac{r}{1+r}\,z\right)}
{1-\frac{r}{1+r}\,z-\frac{z^2}{1+r} \left[1-\tau^{-1-r(1-z)}\right]}.
\end{eqnarray}

The term $\tau^{-1-r(1-z)}$ in the denominator can be ignored
within the boundary layer.  Therefore, the denominator becomes
$(1-z)[1+(1+r)^{-1}z]$, and the generating function simplifies to
\begin{eqnarray}
\label{G2in}
G(z,t)=c_2(t)\,z^2\,e^{\lambda z}\,\frac{1}{2+r}\left[
\frac{1}{1-z}+\frac{1+r}{1+\frac{z}{1+r}}\right]\,.
\end{eqnarray}
To extract the mass distribution, we expand $G(z,t)$ in a Taylor series in
$z$.  This gives
\begin{eqnarray}
\label{ck2in}
\frac{c_{k+2}}{c_2}=\sum_{n=0}^k\frac{\lambda^n}{n!}
\left[\frac{1}{2+r}+\frac{1+r}{2+r}\left(-\frac{1}{1+r}\right)^{k-n}\right]\,.
\end{eqnarray}
Thus apart from the leading contribution that equals $\lambda^k/k!$, in
agreement with our previous result in Eq.~(\ref{ck1}), we find all the
correction terms; {\it e.g.}, the leading correction is
$(1+r)^{-1}\lambda^{k-2}/(k-2)!$.

When both the quadratic and the transcendental terms in the denominator
balance each other, the generating function accounts for the main part of the
mass distribution.  The both terms are comparable when $1-z\propto
\tau^{-1}$.  Writing $1-z=\zeta/\tau$ and taking $z\to 1$, $\tau\to \infty$
limit, with $\zeta$ kept finite, we simplify the generating function:
\begin{eqnarray}
\label{G2out}
G=\frac{c_2(0)}{\tau}\,\frac{1}{1+(2+r)\zeta}\,.
\end{eqnarray}
This form of the generating function implies that
\begin{eqnarray}
\label{ckout}
c_{k}=\frac{c_2(0)}{(2+r)\,\tau^2}\,
\exp\left[-\frac{1}{2+r}\,\frac{k}{\tau}\right]\,.
\end{eqnarray}
Since $\gamma=2$ and $(2+r)R(0)=2c_2(0)+c_1(0)=1$ (for the bi-disperse
initial conditions $R(0)=c_2(0)$ and the mass density is always set equal to
one), the above scaling form reduces to the anticipated scaling form of
Eq.~(\ref{ckscal}).

\end{document}